\documentclass[referee]{raa}           
\usepackage{graphicx,times}
\usepackage{natbib}
\usepackage{amssymb,amsmath}
\bibpunct{(}{)}{;}{a}{}{,}
\usepackage{soul}
\usepackage{xcolor}

\usepackage[pagebackref=true]{hyperref}
\begin{document}

\title{Searching for the highest energy of pulsation and critical luminosity of Swift J0243.6+6124 observed by \textit{Insight}-HXMT}

\volnopage{ {\bf 20XX} Vol.\ {\bf X} No. {\bf XX}, 000--000}
\setcounter{page}{1}

\author{Qing-Xia Zhao
\inst{1,2,3}, Xian Hou\inst{1,4}, Ming-Yu Ge\inst{2}, Shuang-Nan Zhang\inst{2}, Yun-Xiang Xiao\inst{2,3}, You-Li Tuo\inst{5}, Zi-Xu Yang\inst{6}, Ling-Da Kong\inst{5}, Jin-Lu Qu\inst{3}, Shu Zhang\inst{3}, and Jian-Cheng Wang\inst{1,4}}

\institute{ Yunnan Observatories, Chinese Academy of Sciences, Kunming 650216, China; {\it zhaoqingxia@ynao.ac.cn}; {\it xhou@ynao.ac.cn}\\
\and
Key Laboratory of Particle Astrophysics, Institute of High Energy Physics, Chinese Academy of Sciences, Beijing 100049, China; {\it gemy@ihep.ac.cn}\\
\and
University of Chinese Academy of Sciences, Chinese Academy of Sciences, Beijing 100049, China\\
\and
Key Laboratory for the Structure and Evolution of Celestial Objects, Chinese Academy of Sciences, Kunming 650216, China\\
\and
Institut f{\"{u}}r Astronomie und Astrophysik, Sand 1, D-72076 T{\"{u}}bingen, Germany\\
\and
School of Physics and Optoelectronic Engineering, Shandong University of Technology, Zibo 255000, China\\
}
\vs \no
   {\small Received 20XX Month Day; accepted 20XX Month Day}

\abstract{Owing to the broad energy coverage of \textit{Insight}-HXMT in the hard X-ray band, we detected the highest energy of pulsation exceeding 200 keV around the 2017-2018 outburst peak of the first Galactic pulsating ultraluminous X-ray source (PULX) Swift J0243.6+6124, which is the highest energy detected from PULXs to date. We also obtained the highest energy of pulsation of every exposure during the outburst in 2017-2018, and found the highest energy is roughly positively correlated with luminosity. Using our newly developed method, we identified the critical luminosity being $4\times 10^{38}\, \rm erg\,s^{-1}$ when the main peaks of the low and high energy pulse profiles became aligned, which separates the fan-beam dominated and pencil-beam dominated accretion regimes. Above the critical luminosity, the phase of the main peak shifted gradually from 0.5 to 0.8 until the outburst peak in all energy bands is reached, which is in agreement with the phase shift found previously at low energies. Our result is consistent with what is derived from spectral analysis.
}

   \authorrunning{Q.-X. Zhao et al.}            
   \titlerunning{The highest energy of pulsation and critical luminosity of Swift J0243.6+6124}  
   \maketitle

\section{Introduction}           
\label{sect:intro}

BeXRBs are a sub-class of high mass X-ray binaries (HMXBs) consisting of a neutron star (NS) accreting material and angular momentum from a Be companion star. They are among the brightest X-ray sources in the sky. Giant (Type II) outbursts are usually observed from BeXRBs which can exceed an X-ray luminosity of $10^{37}\, \rm erg\ s^{-1}$ and last for several weeks to even months. The giant outbursts provide a good opportunity to study the accretion process and emission properties of BeXRBs \citep{Reig2011}. 

The radiation structure and properties of BeXRBs depend on the luminosity and accretion rate. At lower luminosities, the radiation is produced near the polar caps of the NS and escape mostly along the magnetic axis forming a ``pencil'' beam pattern \citep{Basko1975,Burnard1991,Nelson1993,Becker2012}, while at higher luminosities, materials are decelerated by the radiation-dominated shock and accretion columns are formed at the magnetic poles. In this case, the X-ray emission runs away predominantly through the column wall forming a ``fan'' beam pattern \citep{Davidson1973,Basko1976,Becker2012,Mushtukov2015b}. The transition luminosity between these two accretion regimes is called the ``critical luminosity'' ($L_{\rm crit}$), the detection of which based on the different beam patterns was proposed by \cite{Gnedin1973}. In the radiation model of \cite{Becker2012}, when the luminosity is below $L_{\rm crit}$ (subcritical accretion), pencil beam radiation dominates (mixture of pencil and fan beam), and above $L_{\rm crit}$ (supercritical accretion) fan beam radiation dominates. The observed pulse profiles depend on luminosity, energy and the geometry of the rotating NS in the observer's reference frame. Many studies have shown that reflection by the NS surface and gravitational lensing effects result in more complex beam patterns than the simplified ``pencil-fan'' picture and thus more complex pulse profiles \citep[see e.g.,][]{Poutanen2013,Lutovinov2015,Postnov2015,Mushtukov2018,Markozov2024}, as usually observed.

Determining $L_{\rm crit}$ is not only crucial to identify the emission pattern transition which helps to probe the accretion state of BeXRBs, but also useful to estimate the magnetic field of the NS \citep[see e.g.,][]{Wang1981b,Becker2012,Mushtukov2015b}. However, due to the complexity of the pulse profiles especially at low energies, the determination of $L_{\rm crit}$ using only low energy data (e.g., below about 30 keV) is highly uncertain. \cite{Hou2022} (and more details in the Appendix of this work) demonstrated that the beam pattern, thus $L_{\rm crit}$, can be identified by judging whether the phases of the high and low energy main peaks are aligned in phase, since high energy photons (e.g., $>$50\,keV) preferentially escape from the wall of the column, i.e., in the fan beam pattern. 

Ultraluminous X-ray (ULX) sources are ultra-bright sources with apparent X-ray luminosity greater than $1\times10^{39}\, \rm erg\ s^{-1}$ (the Eddington limit of a $\sim 10M_{\rm \odot}$ black hole). Known as pulsating ULXs (PULXs), some of ULXs are found to have X-ray pulsations, establishing that the central engines are NSs \citep{2014Natur.514..202B,2016ApJ...831L..14F,2017MNRAS.466L..48I,2017Sci...355..817I,2018MNRAS.476L..45C,Chandra2020,2020ApJ...895...60R,Vasilopoulos2020, Weng2017}. 
See \cite{King2023} for a recent review on ULXs. Swift J0243.6+6124 is the first Galactic PULX with a peak luminosity of $>1\times10^{39}\, \rm erg\ s^{-1}$ during its 2017-2018 giant outburst. It has a Be companion star, a pulse period of 9.8 s, an orbital period of 27.7\,days and a distance of 6.8\,kpc \citep[See e.g.,][]{Tsygankov2018,Wilson-Hodge2018,Zhang2019,Doroshenko2020,Kong2020,Kong2022,Wang2020}. 

\textit{Insight}-HXMT has a wide energy band of 1$-$250 keV, and the largest effective area above 80 keV. It is the only satellite that can measure pulse energies exceeding 100 keV and its wide energy range provides the opportunity to determine $L_{\rm crit}$ using the method that is described above. In previous studies, $>$130 keV pulsations have been detected from the PULX RX J0209.6$-$7427 \citep{Hou2022} using \textit{Insight}-HXMT data, and the method has been verified in two accreting X-ray pulsars EXO 2030+375 \citep{Fu2023} and RX J0440.9+4431 \citep{Li2023}, respectively.

In this paper, we search for the highest energy of pulsation of Swift J0243.6+6124 using the broad-band observations by \textit{Insight}-HXMT, and determine $L_{\rm crit}$ using our method. We describe the observation and data analysis in Section \ref{sect:data} and present our results in Section \ref{sect:result}. We finally discuss our findings in Section \ref{sect:discussion} and summarize in Section \ref{sect:summary}.

\section{Observations and data analysis}
\label{sect:data}

\textit{Insight}-HXMT carries three slat-collimated X-ray telescopes: the High Energy X-ray telescope \citep[HE,][]{Liu2020}, the Medium Energy X-ray telescope \citep[ME,][]{Cao2020}, and the Low Energy X-ray telescope \citep[LE,][]{Chen2020}. Its broad energy band (1$-$250\,keV), large detection area (5100 cm$^2$ in 20$-$250\,keV for HE) and the small dead time make it a powerful satellite in X-ray studies of bright X-ray sources \citep{Zhang2020}.

Swift J0243.6+6124 was observed continuously by \textit{Insight}-HXMT from 2017 October 7 (MJD 58033) to 2018 February 21 (MJD 58170) covering its entire 2017-2018 outburst, providing rich data for investigating the properties of this source. {\it Insight}-HXMT made 102 pointing observations of Swift J0243.6+6124 during this outburst. The \textit{Insight}-HXMT Data Analysis Software (\texttt{HXMTDAS}) v2.04 with default filters is used to reduce the data. These include screening data with elevation angle (ELV) $>$ 10$^{\circ}$, geometric cutoff rigidity (COR) $>$ 8\,GeV, offset for the point position $\leq 0\fdg04$, and time beyond 300\,s to the South Atlantic Anomaly. The photon arrival time is corrected by the solar system barycenter ($\rm R.A. = 40\fdg918, \rm decl. = 61\fdg4341$) and by the binary orbit with the orbital parameters from Fermi/GBM measurements\footnote{https://gammaray.msfc.nasa.gov/gbm/science/pulsars/lightcurves/swiftj0243.html}. 

To obtain phase-coherent pulse profiles, we fit spin frequency over time using the cubic spline interpolation, where the frequencies are from Fermi/GBM measurements, and then convert photon time after corrections to a sequential pulse phase $\phi(t)$, where $\phi_{0}$ is set to 0 at $t_{0} $ = 59027.49906951 (MJD), which is the epoch of the first Fermi/GBM periodicity detection \citep[see][for details of this method]{Sugizaki2020}.
In addition, the dead time effect has been corrected for the pulse profiles of HE and ME using the recorded dead time in view of the high luminosity of Swift J0243.6+6124 and the corresponding significant dead time effect.

\section{Results}
\label{sect:result}
\subsection{Search for the highest energy of pulsation}

\begin{figure*}
    \begin{center}
    \includegraphics[width=0.75\linewidth]{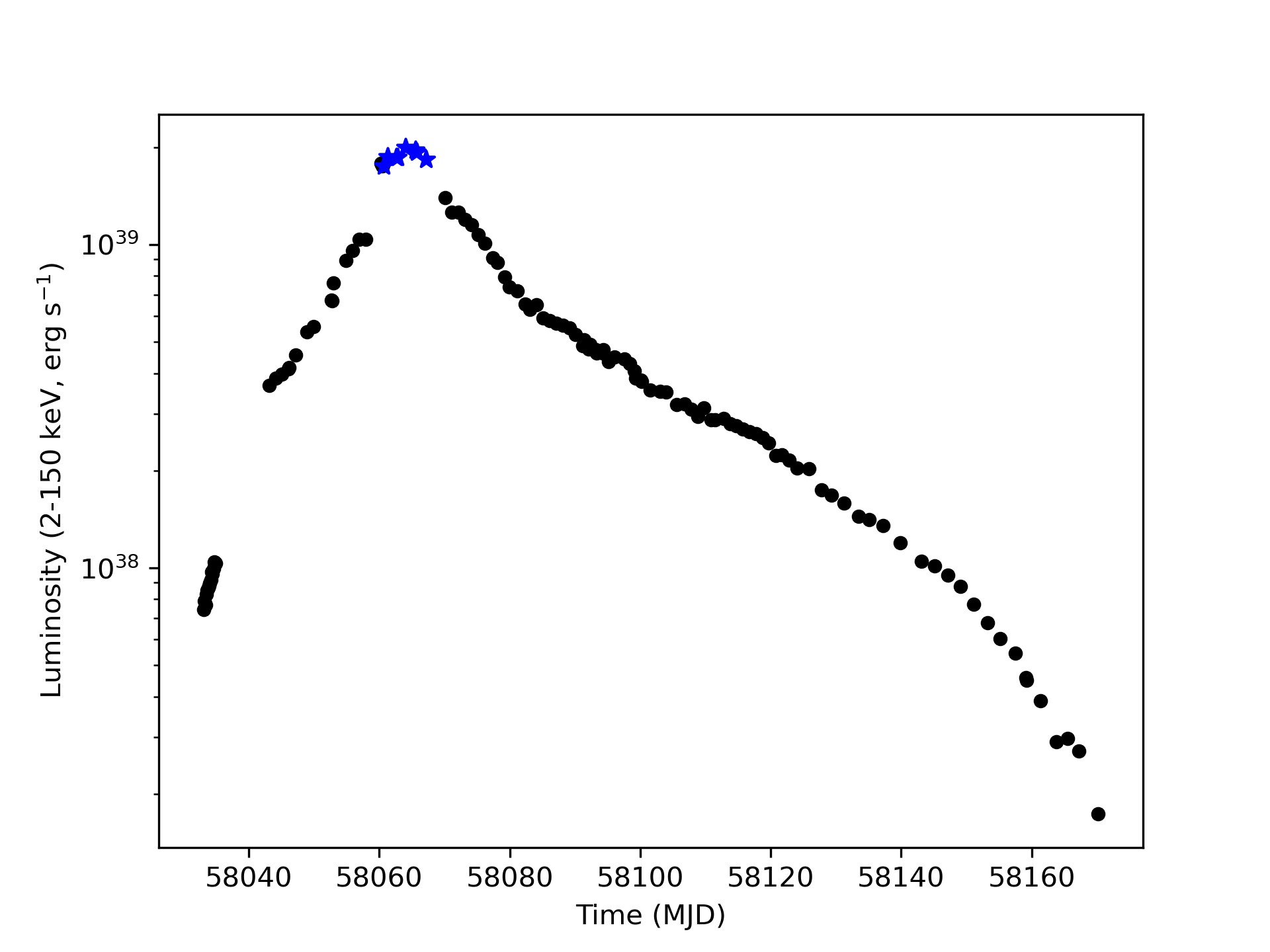}
    \caption{The luminosity evolution of Swift J0243.6+6124 as observed by \textit{Insight}-HXMT from MJD 58033 to 58170 (2017 October 7 to 2018 February 21) in the 2$-$150 keV range. The blue stars mark the nine exposures at the outburst peak.}
    \label{fig:lumin_time}
    \end{center}
\end{figure*}

Figure \ref{fig:lumin_time} shows the luminosity evolution of Swift J0243.6+6124 from MJD 58033 to 58170 (2017 October 7 to 2018 February 21) in the 2$-$150 keV range, and the blue stars mark the nine exposures to be combined at the outburst peak. The luminosity was calculated by multiplying the flux by $4\pi{d}^{2}$, where the distance $d$ was taken as 6.8\,kpc, and the flux was estimated by fitting the broad-band \textit{Insight}-HXMT spectra using \texttt{constant*tbabs*(bbodyrad+gauss+cutoffpl)} model.
In order to search for the highest energy of pulsation, we combined nine exposures at the outburst peak between MJD 58060 and 58068 (2017 November 3 to 2017 November 11, ObsID: P011457701704-P011457702301). The pulse profiles of these nine exposures remained stable, therefore, it is reasonable to combine them. 

\begin{figure*}
    \begin{center}
    \includegraphics[width=0.5\linewidth]{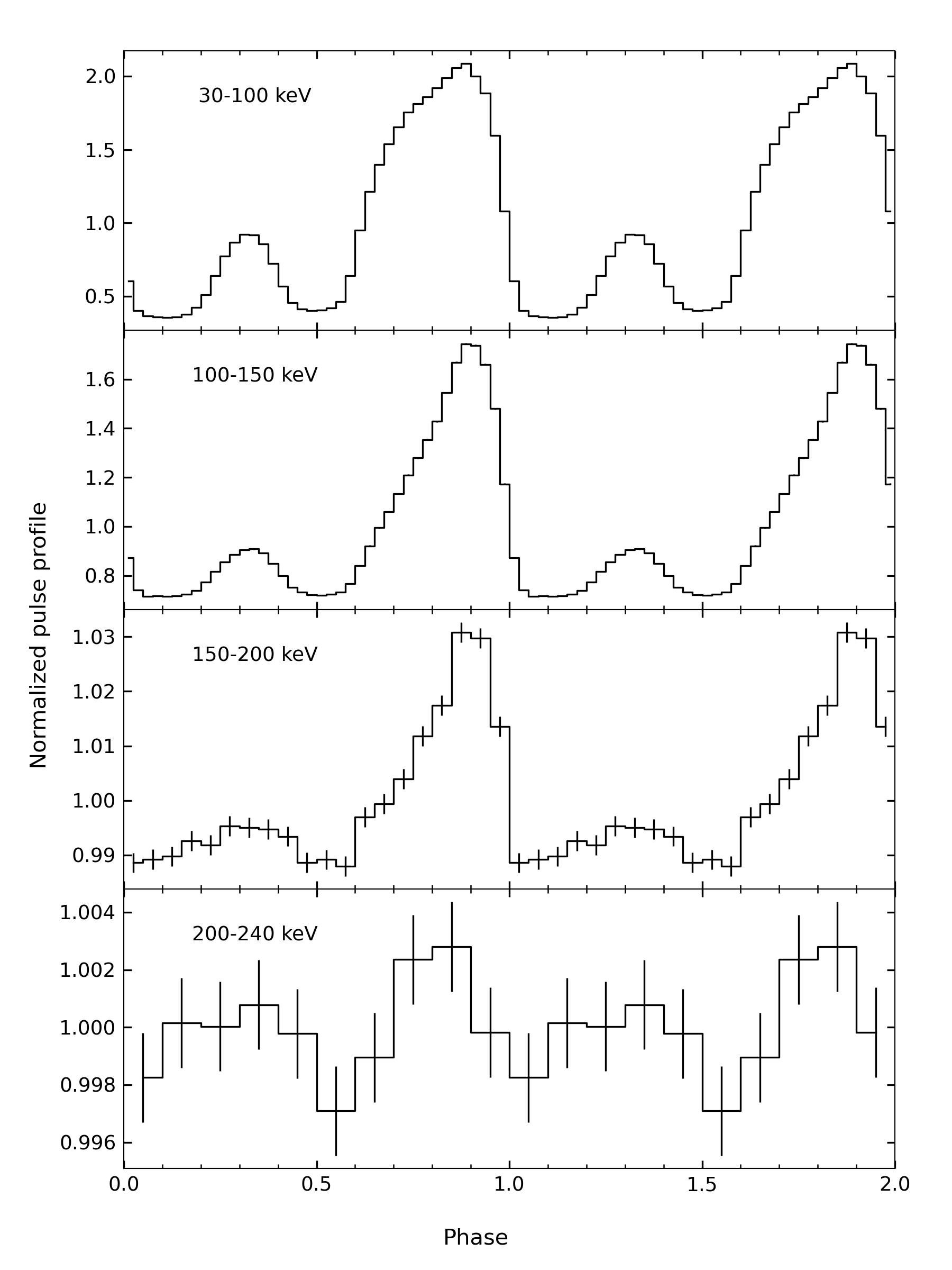}
    \caption{Pulse profiles in different energy bands by combining the nine exposures from 2017 November 3 to November 11 at the outburst peak of Swift J0243.6+6124. The pulse profiles are normalized by the mean count rate.}
    \label{fig:profile_peak}
    \end{center}
\end{figure*}

\begin{figure*}
    \begin{center}
    \includegraphics[width=0.75\linewidth]{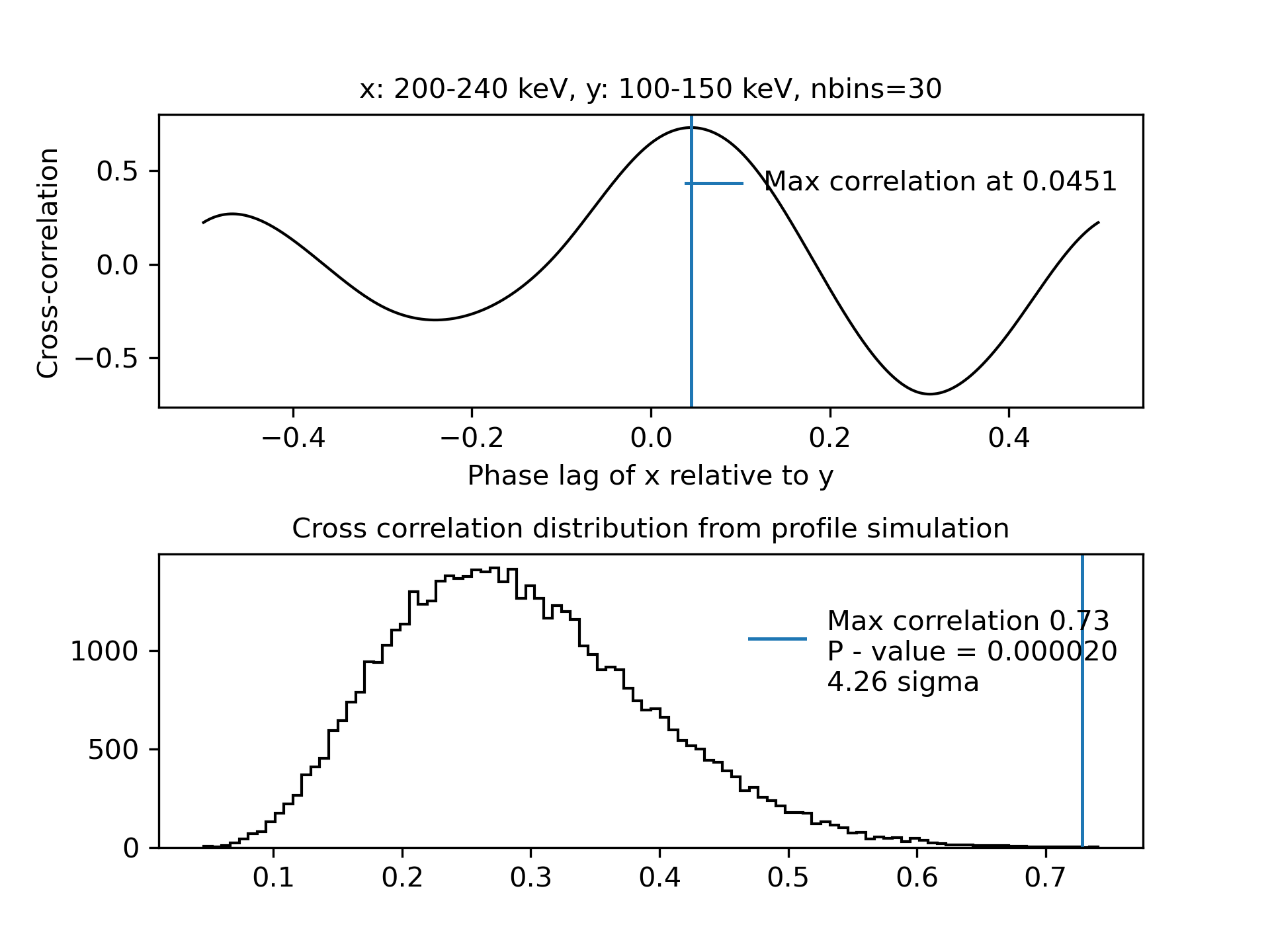}
    \caption{Cross-correlation of the pulse profiles between 200$-$240 keV and 100$-$150 keV. The upper panel shows the cross-correlation  with phase lags. The lower panel shows the cross-correlation distribution from $10^5$ Monte Carlo simulations of two non-correlated profiles.}
    \label{fig:Crosscorr_significance}
    \end{center}
\end{figure*}

When folding pulse profiles using the events after barycentric correction, GTI boundary correction was performed, i.e. only the events in integer periods were adopted to fold the pulse profile. The pulse profiles of the nine exposures were created in different energy bands respectively and combined using the cross-correlation technique to align phases. Figure \ref{fig:profile_peak} shows the combined pulse profiles in different energy bands. We can see that the profile has one main peak and one minor peak which are separated by $\sim$ 0.5 in phase. As the energy increases, both peaks retain until above 200 keV.
The significance of the pulsation is quantified using the cross-correlation method that developed by \cite{Hou2022}. We cross correlated the 200$-$240 keV profile with the profile in the lower energy band of 100$-$150 keV, which has similar profile shape but higher signal-to-noise ratio. The specific method is as follows: First, we calculate the cross correlation of the 200$-$240 keV profile with the 100$-$150 keV profile. Then we use the Monte Carlo method to simulate the cross-correlation distribution between two non-correlated profiles: one is generated by performing Poisson sampling of a flat profile and the mean of the simulated profile is taken as that of the 200$-$240 keV profile; the second profile is generated by performing Poisson sampling for each bin of the 100$-$150 keV
profile. The cross-correlation of these two sampled profiles is then calculated, and finally the simulation is repeated for $10^5$ times to obtain the distribution of the cross correlation. The result is shown in Figure \ref{fig:Crosscorr_significance}.

In the upper panel of Figure \ref{fig:Crosscorr_significance}, the blue line indicates the phase lag at which the maximum cross-correlation value locates. It can be seen that the phase lag of the two profiles is small ($\sim 0.0451$), which means that the profile of the high energy band is in phase with that of the lower energy band. In the lower panel, the blue line indicates the maximum cross-correlation value calculated from the upper panel. Then we evaluate the significance of such a cross correlation by calculating the proportion of the part on the right of the blue line, i.e., the chance probability ($p$-value) that the result is obtained by chance assuming that the null hypothesis (the 200$-$240 keV profile is flat) is true. The significance is 4.26$\sigma$ assuming a normal distribution. So we are confident that the highest energy of the pulsation of Swift J0243.6+6124 exceeds 200 keV, which is the highest energy detected so far from known PULXs, exceeding the 130 keV detected from RX J0209.6-7427 \citep{Hou2022}.

\subsection{Evolution of the highest energy of pulsation }

\begin{figure*}
  \begin{center}
  \includegraphics[width=0.75\linewidth]{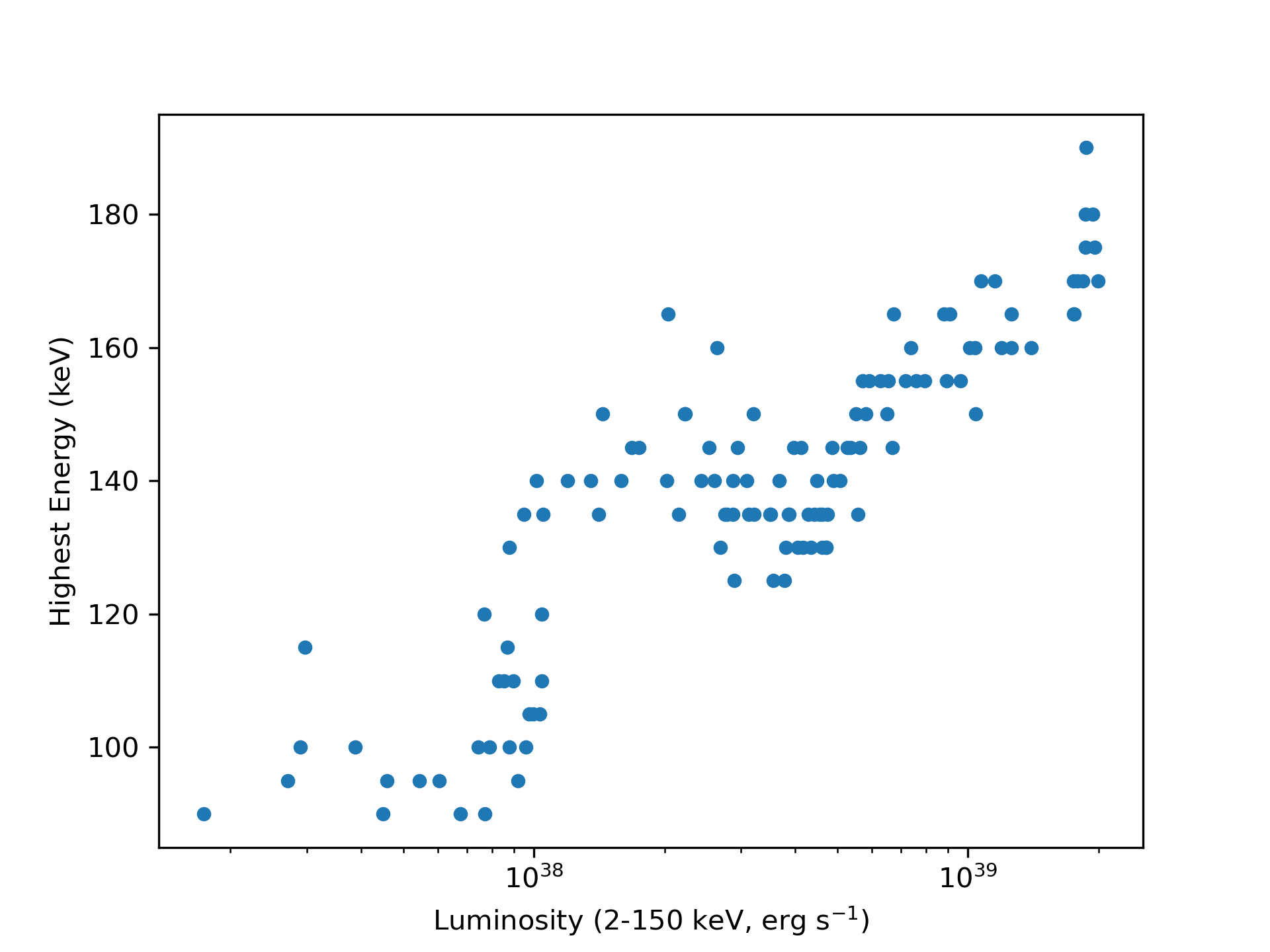}
  \caption{Evolution of the highest energy of pulsation with luminosity for all exposures from MJD 58033 to MJD 58170 covering the whole outburst of Swift J0243.6+6124.}
  \label{fig:HighEn_lumi}
  \end{center}
\end{figure*}

\begin{figure*}
  \begin{center}
  \includegraphics[width=0.9\linewidth]{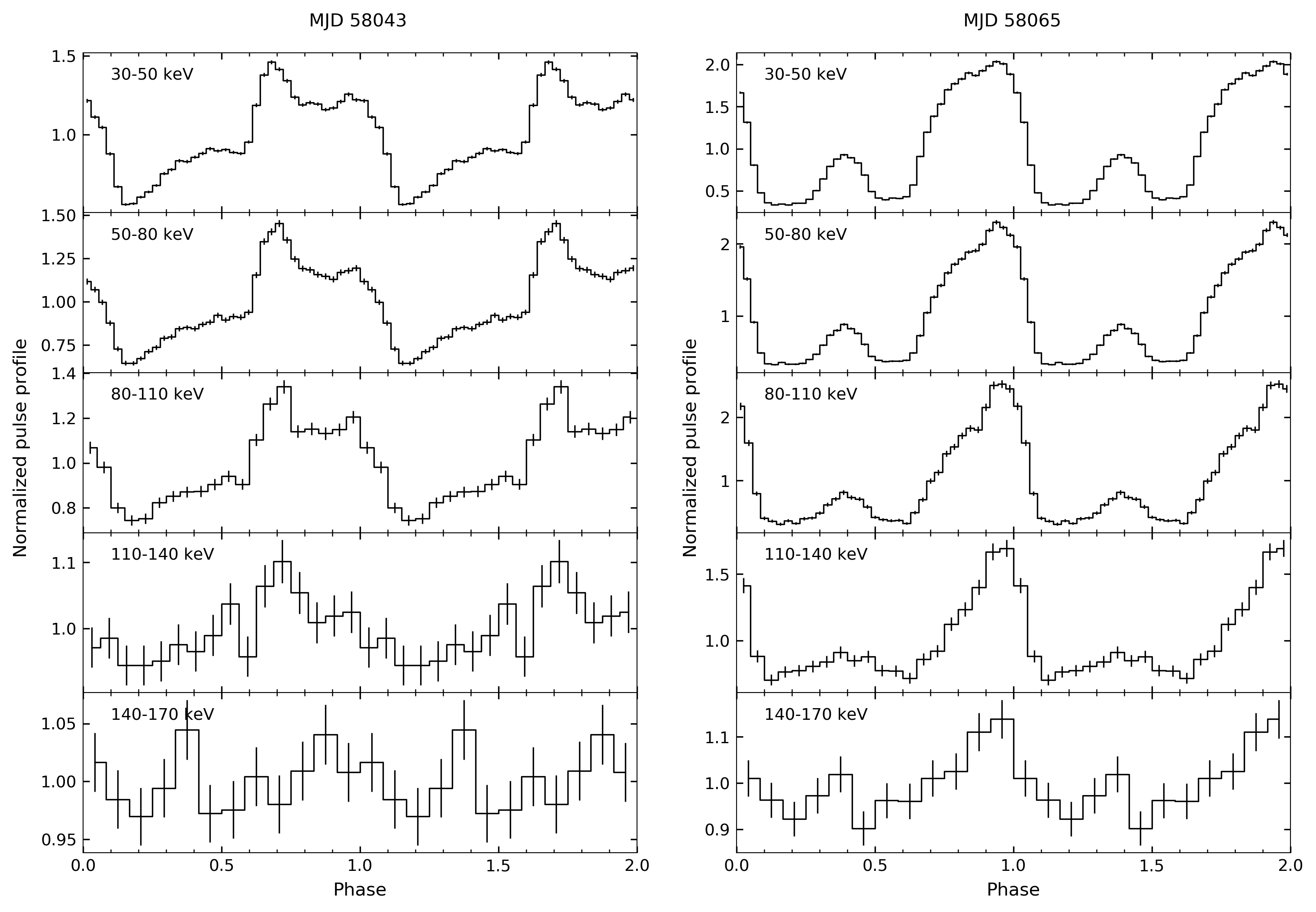}
  \caption{The pulse profiles of two exposures (MJD 58043 and MJD 58065) at luminosities of $3.7\times10^{38}\, \rm erg\ s^{-1}$ (left) and $2\times10^{39}\, \rm erg\ s^{-1}$ (right), respectively, randomly sampled with $10^6$ photons. The highest energy of pulsation of MJD 58043 reaches 110 keV, while that of MJD 58065 reaches 140 keV.}
  \label{fig:profile_sample}
  \end{center}
\end{figure*}

We also searched the highest energy of pulsation of every exposure of the entire outburst from 2017 October 7 (MJD 58033) to 2018 February 21 (MJD 58170) using the method described above, and investigated the evolution of the highest energy of pulsation with luminosity (2$-$150 keV). It appears that the highest energy of pulsation is roughly positively correlated with luminosity, as shown in Figure \ref{fig:HighEn_lumi}. To exclude the possibility that the evolution trend is entirely influenced by photon statistics, we randomly sampled $10^6$ photons for different exposures to find the highest energy of pulsation. Figure \ref{fig:profile_sample} shows the sampled pulse profiles of two exposures at MJD 58043 and MJD 58065 with luminosity of $3.7\times10^{38}\, \rm erg\ s^{-1}$ and $2\times10^{39}\, \rm erg\ s^{-1}$, respectively. The highest energy of pulsation of MJD 58043 reaches 110 keV, while that of MJD 58065 reaches 140 keV. Therefore, we propose that the highest energy of pulsation increases with the increase of the luminosity and such evolution is intrinsic rather than pure statistical effect.

\subsection{Determine the critical luminosity}

We found that once the accretion column is formed at luminosity around (from below to above) $L_{\rm crit}$, high energy photons (e.g., \textgreater 50 keV) preferentially escape from the wall of the column, i.e., in the fan beam pattern. More specifically, they escape from the bottom wall of the column, since the optical depth ratio ($\tau_{\parallel}/\tau_{\perp}$, i.e., parallel/perpendicular to the magnetic field) increases rapidly with energy, and it is thus hard for photons to escape from both the top and the top wall of the accretion column \citep[and more details in the Appendix of this work]{Hou2022}. This means that we can use the highest energy pulse profile to pin-point the spin phase when the radiation from the wall of the accretion column is observed with the maximum flux, i.e., the phase of the fan beam maximum, even at luminosity below $L_{\rm crit}$. The highest energy pulse profile alone cannot be used to identify if the accretion is below or above $L_{\rm crit}$, since the radiation below about 30 keV dominates the total X-ray radiation and thus the radiation at higher energy is only a very small part of the total X-ray luminosity. In other words, we need to find out the transition luminosity when the low energy pulse is dominated by the fan beam emission. However, lower energy photons can contribute to both the pencil and fan beams, though predominately in the pencil beam below $L_{\rm crit}$ or in the fan beam above $L_{\rm crit}$. 

\begin{figure*}
    \begin{center}
    \includegraphics[width=0.75\linewidth]{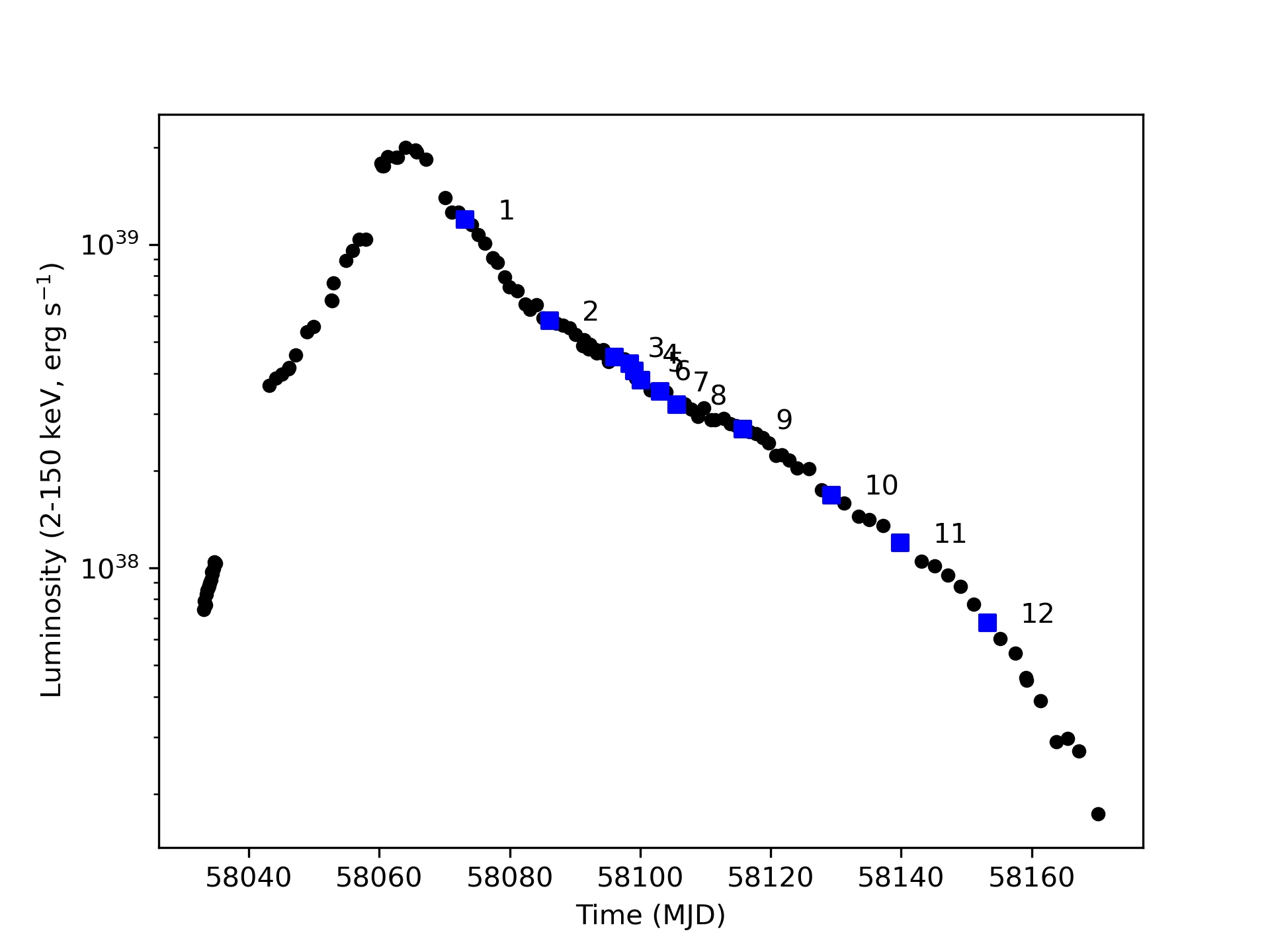}
    \caption{The same light curve of Swift J0243.6+6124 as Figure \ref{fig:lumin_time}. The blue squares mark the exposures the pulse profiles of which are shown in Figure \ref{fig:profile_Lcrit_12}, and the numbers on the right of the squares are serial numbers corresponding to those of the pulse profiles shown in Fig.\ref{fig:profile_Lcrit_12} .}
    \label{fig:lumin_time_12}
    \end{center}
\end{figure*}

Therefore, our strategy is to use the highest energy pulse peak to first pin-point the spin phase when the radiation from the wall of the accretion column is observed with the maximum flux, i.e., to find the peak phase of the highest energy emission, and then examine at what luminosity the low energy pulse peak becomes aligned with the high energy pulse peak of the fan beam pattern. This luminosity should be $L_{\rm crit}$.
We thus generated the pulse profiles evolving with energy for every single exposure.
In order to cover a wide range of luminosity, we selected 12 exposures during the outburst decay as indicated in Figure \ref{fig:lumin_time_12}, and presented their pulse profiles evolving with energy in Figure \ref{fig:profile_Lcrit_12}. 
According to the pulse profile evolution from Panels (12) to (5), i.e., with the increasing luminosity, the main peak in the high energy band of 50$-$100\,keV remains at around phase of 0.5 (red dotted line). In the pulse profiles of Panel (7) for MJD 58102 and after, the main peaks in high and low energy bands are not in phase, indicating that it is the subcritical regime with pencil beam emission dominating. Then in Panel (5) for MJD 58099, the main peak in the lower energy band (1$-$4\,keV) is also at the same phase of 0.5, i.e., the main peaks in high and lower energy bands are in phase. So, this epoch is in the supercritical regime with fan beam emission dominating. Thus, we propose that the transition from pencil-beam dominated to fan-beam dominated emission regimes occurred at the epoch of Panel (5), and $L_{\rm crit}$ is calculated correspondingly as $4\times10^{38}\, \rm erg\ s^{-1}$.

\section{Discussion}
\label{sect:discussion}
Several attempts have been carried out to determine $L_{\rm crit}$ basing on timing and spectral properties of Swift J0243.6+0124 during its 2017-2018 outburst. Its pulse profiles show complex evolution with significant luminosity and energy dependence. 
Using \textit{NICER} and \textit{Fermi}/GBM observations, 
\cite{Wilson-Hodge2018} showed that near a luminosity of $1\times10^{38}\, \rm erg\ s^{-1}$, the pulse profiles evolved from single-peaked to double-peaked, the pulsed fraction changed from decreasing to increasing with increasing luminosity, and the hardness ratios became softening above $1\times10^{38}\, \rm erg\ s^{-1}$.
These behaviors repeated at the same luminosity during the outburst rising and decaying parts, and this luminosity was interpreted as the $L_{\rm crit}$ which represents the transition of two accretion regimes between a Coulomb collisional stopping mechanism and a radiation-dominated stopping mechanism. 
At about the same luminosity of $1.5\times10^{38}\, \rm erg\ s^{-1}$, \cite{Doroshenko2020} also found the change of pulse profile shape using data from \textit{Insight}-HXMT, and also considered it as a transition between two accretion regimes. 
Besides, they found at higher luminosity of about $4.4\times10^{38}\, \rm erg\ s^{-1}$, there was a change in both pulse profile shape and power spectrum, which they associated with the transition of the accretion disk from a gas pressure dominated state to radiation pressure dominated state. 

However, \cite{Kong2020} considered $4.4\times10^{38}\, \rm erg\ s^{-1}$ as $L_{\rm crit}$ instead at which the accretion regimes transferred based on the observed spectral property that the photon indices changed at this luminosity and above it had positive correlation with luminosity. 
\cite{Kong2022} obtained another value of $L_{\rm crit}$ as of $2.9\times10^{38}\, \rm erg\ s^{-1}$ using the relation $L_{\rm crit} = 1.5 \times L_{37}B_{12}^{16/15} $ \citep{Becker2012}, where the NS magnetic field $B$ was estimated based on the detection of the cyclotron resonance scattering feature (CRSF) around 120$-$146 keV from this source.
\cite{Liu2022} found there was a 0.25 phase offset between the double-peak profiles at luminosities below $1.5\times10^{38}\, \rm erg\ s^{-1}$ and above $5\times10^{38}\, \rm erg\ s^{-1}$, and considered $1.5\times10^{38}\, \rm erg\ s^{-1}$ as $L_{\rm crit}$ and the onset of the accretion column.
The critical luminosity of $4\times10^{38}\, \rm erg\ s^{-1}$ determined from our method in this work is close to that obtained by \cite{Kong2020} based on the evolution of photon indices with luminosity. 

\begin{figure*}
    \begin{center}
    \includegraphics[width=1.0\textwidth]{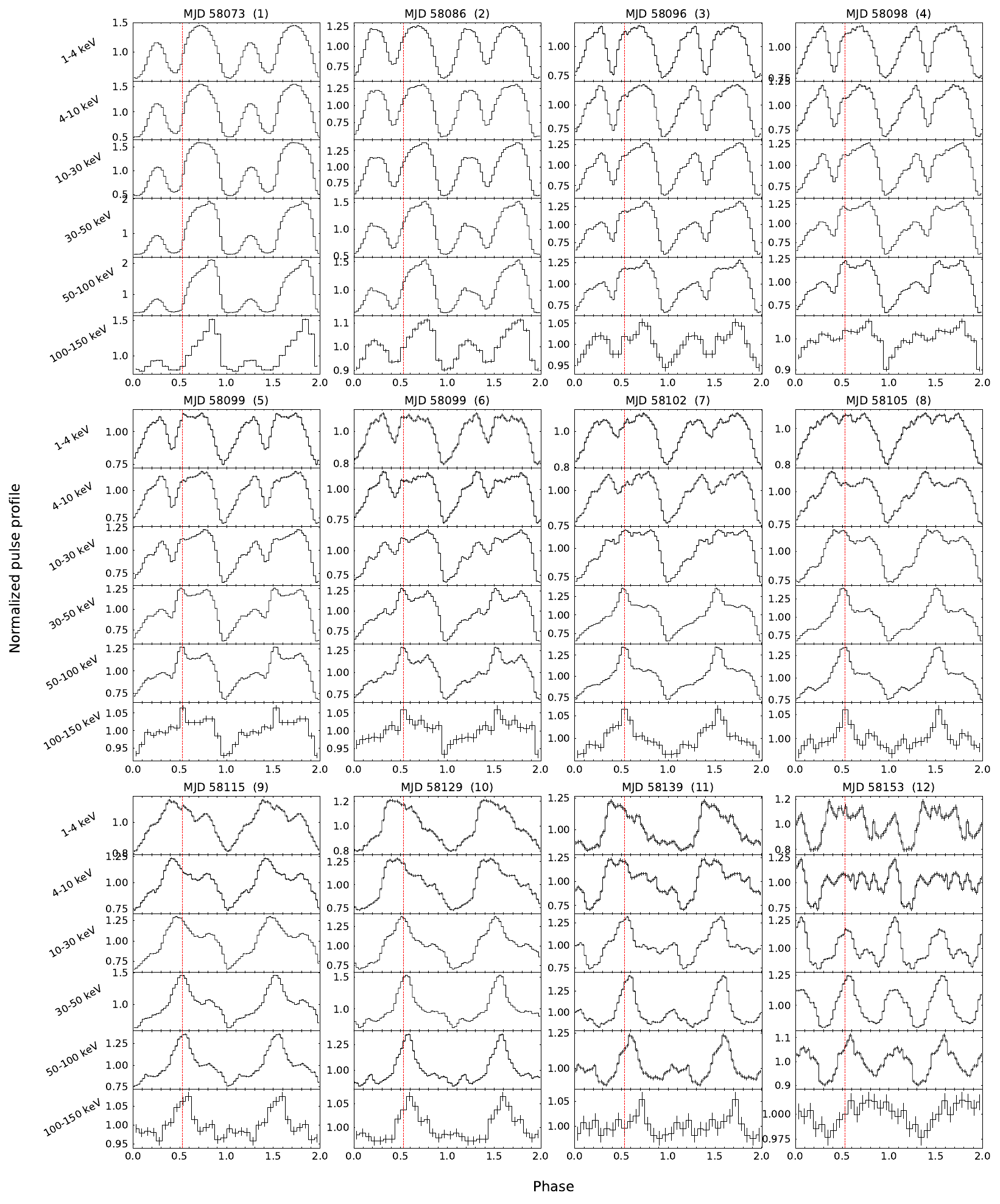}
    \caption{Pulse profiles of Swift J0243.6+6124 evolving with energy of 12 exposures during the outburst decay. The red dotted lines in Panel (5) mark the phase of the main peak in high energy. In other panels, lines are marked at the same phase as those in Panel (5).}
    \label{fig:profile_Lcrit_12}
    \end{center}
\end{figure*}

From Figure \ref{fig:profile_Lcrit_12}, we can see at even higher luminosities (Panels (1)-(4)) than $L_{\rm crit}$ (Panel (5)), the main peak in the high energy band shifted from phase 0.5 to around 0.8. The phase shift was also noted in \cite{Doroshenko2020}, but was obtained based on lower energy pulse profiles and they did not show the case for the pulsation at higher energies, which is indeed our new result. We note that at such high luminosities, the accretion flow covering the magnetosphere of an NS becomes optically thick. This can strongly influence the formation of pulse profiles resulting in, for example, increased pulse fraction and smooth profiles \citep{Mushtukov2017,Mushtukov2019,Brice2023}, the latter is indeed observed in our analysis (e.g., Panels (1)-(2)).

This phase shift may be explained by the gravitational lensing effect which caused light bending for the photons from the fan-beam of the opposite accretion column. In general, with the growing of the accretion column, i.e., increasing luminosity, we will see more and more photons originating from the opposite accretion column and both the low and high energy profiles will be dominated by the fan-beam emission pattern from both the facing and opposite accretion columns. Detailed discussion of the pulse profile evolution far above $L_{\rm crit}$, by taking into account of the gravitational lensing effect and the accretion column geometry, is out of the scope of this work. 

\section{Summary}
\label{sect:summary}

In summary, we analyzed the pulse profile of Swift J0243.6+0124 based on observations from \textit{Insight}-HXMT, and obtained the highest energy of pulsation and the critical luminosity. The highest energy of pulsation reached above 200 keV, the highest observed so far from pulsating ultraluminous X-ray pulsars. The critical luminosity was determined using our method which involves judging whether the main peaks of the pulse profiles in low and high energy bands are in phase. This method has also been applied successfully to two X-ray binary sources EXO 2030+375 and RX J0440.9+4431. The critical luminosity that we obtained was about $4\times10^{38}\, \rm erg\, s^{-1}$, roughly consistent with that obtained from spectral analysis in previous works.

\normalem
\begin{acknowledgements}
This work used data from the \textit{Insight}-HXMT mission, a project funded by China National Space Administration (CNSA) and the Chinese Academy of Sciences (CAS). This work is supported by the National Key R\&D Program of China (2021YFA0718500, 2023YFE0101200) from the Minister of Science and Technology of China (MOST). The authors thank supports from the National Natural Science Foundation of China under Grants 12373051 and 12333007. This work is also supported by International Partnership Program of Chinese Academy of Sciences (Grant No.113111KYSB20190020).

\end{acknowledgements}

\clearpage


\clearpage

\clearpage

\appendix

\section*{Appendix A}
\setcounter{equation}{0}
\renewcommand\theequation{A\arabic{equation}} 
The critical luminosity distinguishes which one dominates between pencil-beam and fan-beam radiation patterns. The pencil beam is parallel to the magnetic field axis, while the fan beam is perpendicular to the magnetic field. The optical depth determines whether a photon can escape from the accretion column, and is defined as \citep{Arons1987}:
\begin{equation}
\tau = n_e\sigma l,
\end{equation}
where $n_e$ is the electron number density, $\sigma $ is the scattering cross section between photons and electrons, and $l$ is the photon propagation length. It is theoretically possible to determine which radiation pattern dominates by comparing the optical depth parallel to and perpendicular to the magnetic field.
Here we used the simplified Compton scattering cross sections, while the scattering cross sections in a strong field limit were calculated accurately by various authors \citep{Herold1979,Daugherty1986,Mushtukov2016}. The simplified cross sections depend on the photon energy, the magnetic field, the polarization mode, and the photon direction momentum:
\begin{enumerate}
    \item $E < E_c$
    \begin{itemize}
        \item[*] O-Mode
            \begin{equation}
            \begin{aligned}
            \sigma_{\parallel}=\sigma_{\rm T}(E/E_{\rm c})^2,\,\,\\
            \sigma_{\perp}=\sigma_{\rm T}, \, \\
            \end{aligned}
            \end{equation}
        \item[*] X-Mode
            \begin{equation}
            \sigma_{X}=\sigma_{\rm T}(E/E_{\rm c})^2,\,\,\\ \end{equation}
    \end{itemize}
    \item $E \geq E_c$
    \begin{equation}
    \sigma_{\rm O} = \sigma_{\rm X} = \sigma_{\rm T},
    \end{equation}
\end{enumerate}
where the O-mode is the ordinary polarization mode, and its electric vector is in the plane composed of the magnetic field $\boldsymbol{B}$ and the photon momentum $\hbar\boldsymbol{k}$, the X-mode is the extraordinary polarization mode in which the electric vector is perpendicular to the plane of magnetic field and photon momentum, $\sigma_{\rm T}$ is the Thomson scattering cross section, and $E_{\rm c} = 11.6B_{\rm 12}/(1 + z)$ is the CRSF line energy, where $B_{\rm 12}$ is the magnetic field in units of $10^{12}$ G. $n_e$ and $l$ are calculated using the analytical expressions in \cite{Becker2012}. We consider the case where there is radiation-dominated shock such that the accretion column is formed. $n_e$ can be derived from the mass conservation relation, $\dot{M} = \pi r_0^{2}n_e m_p v$, where $v$ equals to the post-shock velocity of the infalling matter, which is 1/7 of the freefall velocity approaching the top of the shock, and $ r_0$ is the radius of the accretion column. $\dot{M}$ is the accretion rate and is related to the X-ray luminosity as $L=GM\dot{M}/R$.

The photon propagation length $l$ in the direction parallel to the magnetic field is estimated by the height of the radiation-dominated shock: 
\begin{equation}
H=1.14\times 10^{5}\,  {\rm cm}\, \left( \frac{M}{1.4\,M_{\rm \odot}}  \right)^{-1} \left( \frac{R}{10\, \rm km} \right) \left(\frac{L}{10^{37}\, \rm erg\, s^{-1}}  \right)   \, , 
\end{equation}
and in the direction perpendicular to the magnetic field $l$ is estimated by the accretion column radius $r_0$ which is the maximum length in this direction: 
\begin{equation}
r_0=1.93 \times 10^5 {\rm cm}\,  \left(\frac{\Lambda}{0.1}\right)^{-1/2} \left(\frac{M}{1.4 \, M_\odot} \right )^{-1/14} \left (\frac{R}{10\, {\rm km}} \right)^{11/14} \left(\frac{B}{10^{12}\, {\rm G}}\right)^{-2/7} \left(\frac{L}{10^{37} \, {\rm erg\, s}^{-1}} \right)^{1/7} .
\end{equation}
In the calculation, we assume typical NS parameters: mass $M = 1.4 M_{\odot}$, radius $R = 10$ km, and redshift $z = 0$. 

For O-mode photons, we calculated the ratio optical depth of parallel and perpendicular to the magnetic field ($\tau_{\parallel}/\tau_{\perp}$) for different photon energies, luminosities and magnetic field strengths, as shown in Appendix Figure \ref{fig:ratio}. We can see that on the one hand, the ratio of optical depth increases rapidly with the increase of photon energy under different luminosities and magnetic field strengths, which suggests that the higher the energy, the more likely the photons escape from the wall of the accretion column in the form of a fan beam. On the other hand, the higher the luminosity, the larger the ratio, i.e., the easier the fan beam is dominated.

\begin{figure}
\renewcommand{\thefigure}{A\arabic{figure}}
\centering
\includegraphics[scale=0.6]{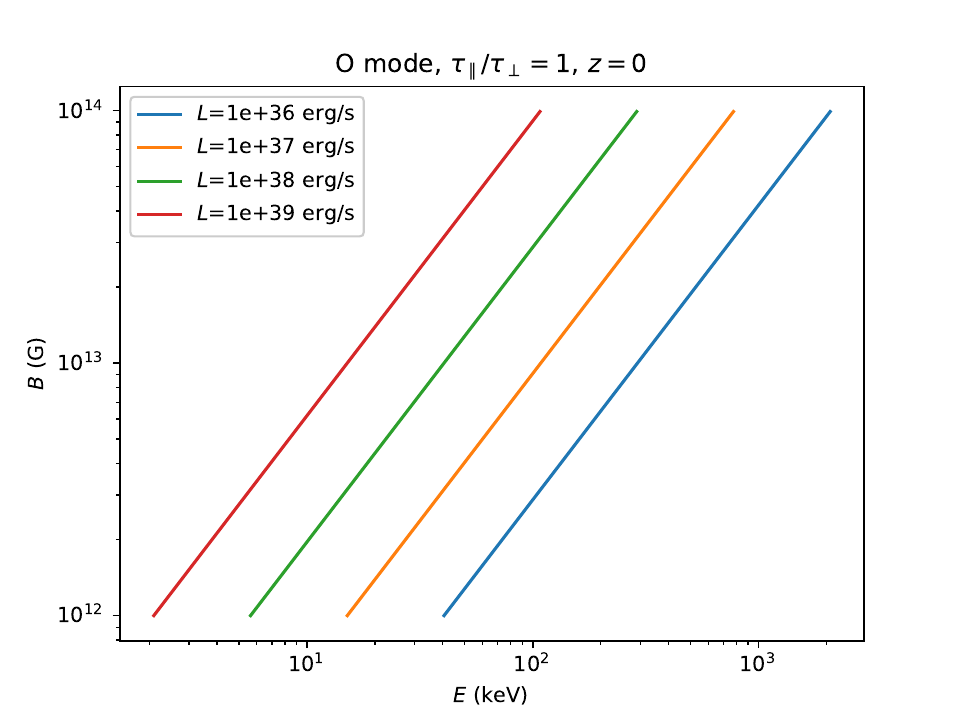}
\caption{Assuming O-mode photons with red shift of 0, a NS with mass of 1.4$M_{\rm \odot}$ and radius of 10 km, the lines show the relation between the magnetic field and energy for different luminosities when the ratio of $\tau_{\parallel}$ to $\tau_{\perp}$ is unity.}
\label{fig:BvsE}
\end{figure}

In addition, Appendix Figure \ref{fig:BvsE} shows the relation of the energies with the magnetic field strengths under different luminosities when the ratio equals unity. This defines the starting energy of the fan beam dominated pattern. We can draw two conclusions: (1) For a given luminosity, the stronger the magnetic field, the higher the starting energy of the fan beam dominated pattern; after crossing the starting energy, the higher the energy, the larger the ratio, the more that is dominated by the fan beam. (2) The higher the luminosity, the lower the starting energy of the fan beam dominated pattern, or the larger the energy range of the fan beam dominated pattern, thus the easier it is to find the moment when the high and low energy pulses begin to be dominated by the fan beam pattern, i.e., they are in phase, and the luminosity at this moment will correspond to $L_{\rm crit}$. For X-mode photons, the scattering cross section is the same in all directions, such that the ratio of the optical depths only depends on the path ratio in both directions and is independent of energy. Therefore, when the actual radiation is a mixture of photons of the two modes, the above conclusions are qualitatively unchanged.

\begin{figure}
\renewcommand{\thefigure}{A\arabic{figure}}
\centering
\includegraphics[scale=0.45]{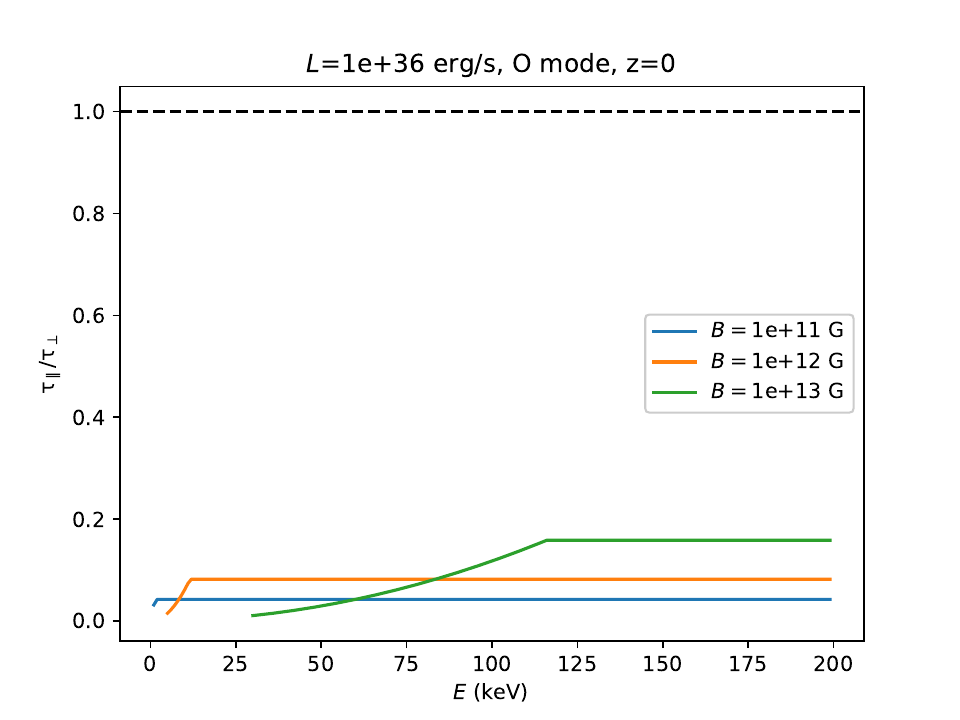}
\includegraphics[scale=0.45]{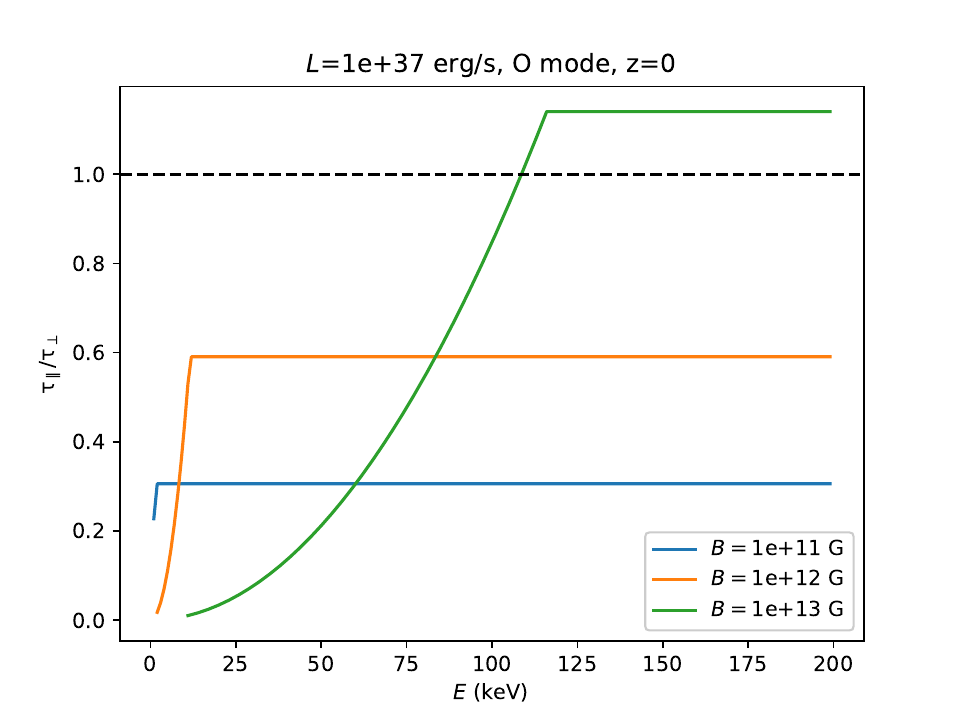}
\includegraphics[scale=0.45]{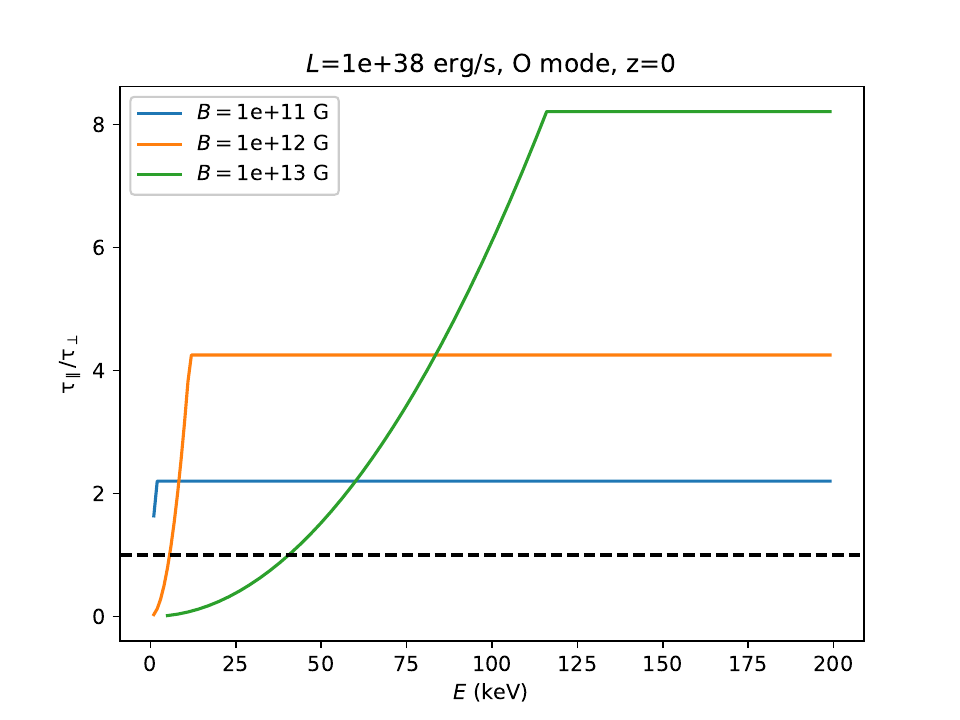}
\includegraphics[scale=0.45]{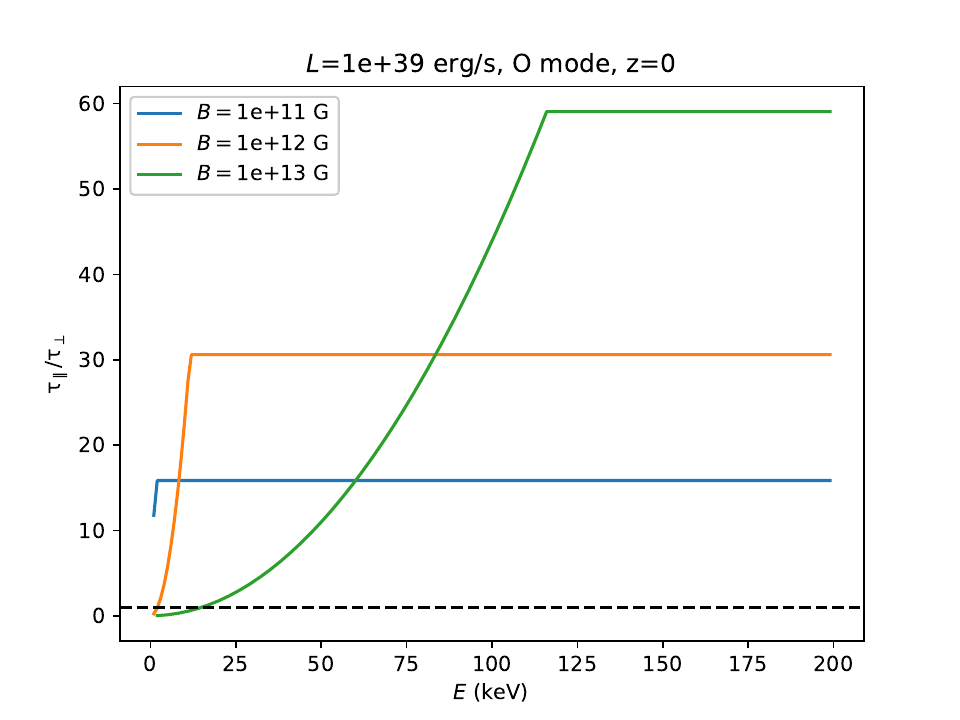}
\caption{Assuming O-mode photons with red shift of 0, an NS with mass of 1.4$M_{\rm \odot}$ and radius of 10 km, the ratio of $\tau_{\parallel}$ to $\tau_{\perp}$ evolves with energy, magnetic field and luminosity. Only the cases with the ratio larger than 0.1 are shown.}
\label{fig:ratio}
\end{figure}

\clearpage

\bibliographystyle{raa}
\bibliography{J0243.bib}

\end{document}